\documentclass[apjl]{emulateapj}

\newcommand{\sups}[1]{\ensuremath{^{\textrm{\scriptsize{#1}}}}} 
\newcommand{\subs}[1]{\ensuremath{_{\textrm{\scriptsize{#1}}}}} 

\newcommand{\sauthor}[2][]{\author{#2\sups{#1}}} 
\newcommand{\saffil}[2][]{\affil{\sups{#1}#2}}

\def\therefore{
\leavevmode
\lower0.1ex\hbox{$\cdot$}
\kern-0.1em\raise0.7ex\hbox{$\cdot$}
\kern-0.1em\lower0.1ex\hbox{$\cdot$}
\;} 

\usepackage{color} 
\usepackage{microtype}
\usepackage{natbib}
\usepackage{graphicx}

\usepackage{float} 

\newcommand{\icarus}{Icarus}

\newcommand{\eqref}[1]{(\ref{#1})}

\newcommand{\cone}{\color{black}}

\begin{document}

\title{Hiding in the Shadows: Searching for Planets in Pre--transitional and Transitional Disks}

\sauthor[1,2]{Jack Dobinson}
\sauthor[1]{Zo\"{e} M. Leinhardt}
\sauthor[3]{Sarah E. Dodson-Robinson}
\sauthor[2]{Nick A. Teanby}
\saffil[1]{School of Physics, H.~H. Wills Physics Laboratory, University of Bristol, Bristol, BS8 1TL, UK}
\saffil[2]{School of Earth Sciences, Wills Memorial Building, University of Bristol, Bristol, BS8 1RJ, UK}
\saffil[3]{Astronomy Department, University of Texas at Austin, Austin, TX 78712, USA}
\slugcomment{Accepted to ApJL September 25, 2013}
\begin{abstract}
Transitional and pre--transitional disks can be explained by a number of mechanisms. This work aims to find a single observationally detectable marker that would imply a planetary origin for the gap and, therefore, indirectly indicate the presence of a young planet. N-body simulations were conducted to investigate the effect of an embedded planet {\cone of one Jupiter mass} on the production of instantaneous collisional dust derived from a background planetesimal disk. Our new model allows us to predict the dust distribution and resulting observable markers with greater accuracy than previous work. Dynamical influences from a planet on a circular orbit are shown to enhance dust production in the disk interior and exterior to the planet orbit while removing planetesimals from the the orbit itself creating a clearly defined gap.
In the case of an eccentric planet the gap opened by the planet is not as clear as the circular case but there is a detectable asymmetry in the dust disk.

\end{abstract}


\maketitle

\section{Introduction}

Transitional and pre-transitional disks are characterised by a flux deficit in the near-infared wavelengths with respect to a classic protoplanetary disk but otherwise have a similar spectral energy distribution (SED). The accepted explanation for the deficit is a lack of optically thick material in part or all of the inner accretion disk \citep{Strom89, Calvet05}. Pre-transitional disks have a gap in the accretion disk with hot optically thick material on the inside, while transitional disks have a complete hole and no detectable hot component to the SED \citep{Espaillat07}.

The origin of the gap or hole in transitional objects is debated and many hypotheses have been suggested, such as particle growth \citep{Dullemond05}, photoevaporation \citep{Alexander07}, and magnetorotational instability \citep{Chiang07}.
However, none of these processes can create the largest holes and gaps observed such as those observed in GM Aur, DM Tau, and SAO 206462, which are tens of AU in radius \citep{Calvet05, Hughes09, Mayama12, Muto12}. This leads us to another very interesting hypothesis for the origin of the interuption in the protoplanetary accretion disk, namely, embedded planetary object(s). The largest holes and gaps require multiple embedded giant planets to explain their extreme size \citep{Rice06, Papaloizou07,Dodson-Robinson11} but many other less extreme transitional objects are consistent with one embedded, unobserved giant planet \citep{Owen11}.

Transitional systems are all young ($\le10$ Myr), thus, if there is an undetected giant planet in the system there should also be a background planetesimal population. The orbits of the background population of planetesimals will be perturbed by the planet \citep{Charnoz01}, increasing collision frequency in regions close to the embedded planet. Dust created will be visible as long as the planetesimals remain collisionally active. This second--generation dust will augment any primordial dust in both the outer gas--rich disk and the inner gas--poor hole by some fraction that is dependent on the survival time of second generation dust grains.
In any disk where planet formation is in progress, second--generation dust will significantly enhance the dust--grain population.

In this Letter we present a search for observational signatures of dust produced by planetesimals under the influence of a giant planet. We run two sets of N--body simulations of a giant planet embedded in a planetesimal population. In one simulation set the collision model (\emph{rubble}) self--consistently treats debris production and in the other set (\emph{perfect merging}) all planetesimal collisions are inelastic, with an analytical prescription for the size distribution of dust particles. We then use a radiative transfer code to create synthetic submillimeter images of the dust disk and identify features that change based on the planet orbit.

\section{Numerical Methods}	
\label{sec:num_methods}

Numerical simulations presented here were calculated using the parallelised N-body gravity code \emph{PKDGRAV} \citep{Richardson00, Stadel01, Leinhardt05}. PKDGRAV uses a hierarchical tree to calculate inter-particle gravity and a second order leapfrog integrator for time evolution. Each simulation took between three and six months to complete on 32 2.8 GHz processors.

\subsection{Initial Conditions and Simulation Details}
\label{sec:IC's+sim_details}
\begin{figure*}[t]
	\includegraphics[width=1.0\textwidth]{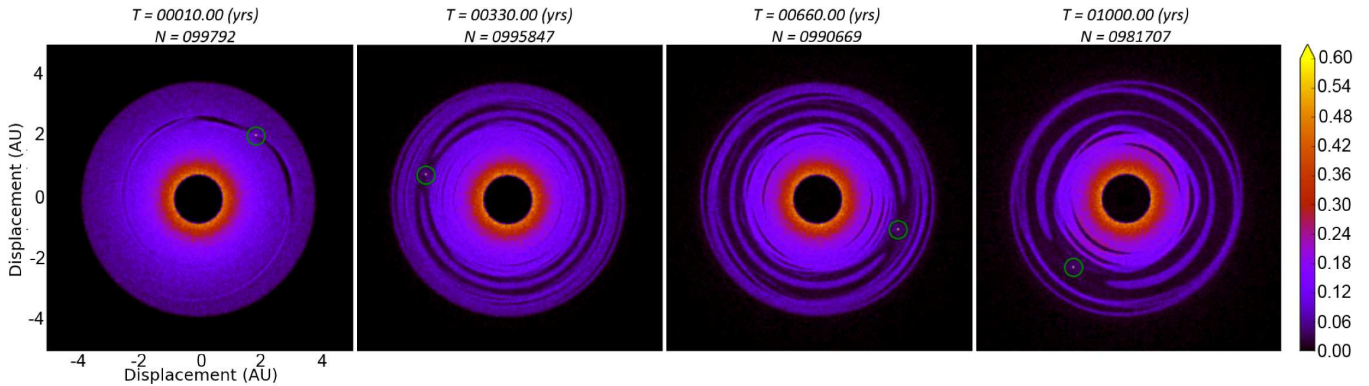}
	\caption{Time evolution of surface density for sim.~4. Pixel area is equivalent to 0.0025 AU\sups{2}. Green circles indicate the position of the growing planet. The colour indicates surface density in mass per pixel area ($M_\oplus/0.0025$ AU$^2$). Time increases left to right.}
	\label{fig:e0_0_merge_surfD_evo}
\end{figure*}

Most simulations begin with $10^6$ equal-mass planetesimals that interact with each other through gravitational forces and physical collisions. Ten simulations were performed in total, in simulations 0-8 $10^6$ planetesimals are initially placed in a 3AU-wide ring from 0.8 to 3.8 AU {\cone around a 1 $M_\odot$ potential}. Simulation 9 has lower particle resolution ($N=10^5$) but a much wider initial disk extending from 0.8 to 10 AU. The embedded planet is placed at 2.8 AU, highlighted by a green circle in Fig.~\ref{fig:e0_0_merge_surfD_evo}.
The total mass of planetesimals is 4.7 M\subs{$\oplus$} {\cone\citep{Richardson00}} with a bulk density of 2 g cm$^{-3}$ \citep{Weidenschilling77}. Planetesimals are initially distributed such that the surface density decreases with semi-major axis, $\Sigma(r) \propto r^{-1.5}$ \citep{Kokubo02,Leinhardt05}.
Eccentricities and inclinations were drawn from a Rayleigh distribution with dispersions $\langle e^2 \rangle = 2\langle i^2\rangle = 0.007$ \citep{Ida92}. Table \ref{table:simic} gives individual simulation details including the initial embedded planet mass, $M_{pl}(0)$, the planetary growth time from $M_{pl}(0)$ to one Jupiter-mass ($M_J$), $T_{grow}$, the eccentricity of the planet, $e_{pl}$, and the final stop time in each simulation, $T_{final}$.

\begin{table}[ht]
\tabcolsep=0.11cm
\small
\caption{Numerical Simulation Initial Conditions}
\centering
\begin{tabular}{c c c c c c}
\hline\hline
Sim & Collision Model & $M_{pl}(0)$ & $T_{grow}$ [yr] & $e_{pl}$ & $T_{final} [yr]$\\
\hline
0 & merging & 1 $M_J$ & 0 & 0.0 & 1000\\
1 & merging & 1 $M_J$ & 0 & 0.1 & 1000\\
2 & merging & 1 $M_J$ & 0 & 0.2 & 1000\\
3 & merging & 1 $M_J$ & 0 & 0.3 & 1000\\
4 & merging & 15 $M_\oplus$ & 1000 & 0.0 & 1000\\
5 & merging & 15 $M_\oplus$ & 1000 & 0.1 & 1000\\
6 & merging & 15 $M_\oplus$ & 1000 & 0.2 & 890\\
7 & rubble & 15 $M_\oplus$ & 1000 & 0.0 & 824\\
8 & rubble & 1 $M_\oplus$ & 1100 & 0.0 & 933\\
9\footnote{\cone the disk in this simulation extends from 0.8 to 10 AU} & merging & 15 $M_\oplus$ & 460 & 0.0 & 1000\\
\hline
\end{tabular}
\label{table:simic}
\end{table}

Computational constraints placed a practical limit of 1000 yr on the simulations. We investigated two initial mass scenarios for the embedded planet: 1) Since the final stage of gas-giant planet growth is short \citep[$\sim 10^3 - 10^4$ yrs,][]{Rice03, Dodson-Robinson08} the planetesimal disk would have little time to respond to the gravitational influence of the planet. Thus, in sim.~0--3 we began with a fully grown $1 M_J$ planet; 2) To simulate more realistic scattering from a growing giant planet we also ran several simulations with an embedded planet that increased in mass exponentially from a terrestrial core to $1 M_J$ over $\sim10^3$ years (sim.~4--8), and 460 years (sim.~9). To grow the planet from a terrestrial-mass embryo to $M_J$
\begin{equation}
dm = \lambda M(t)\, dt, 
\end{equation}
was added at each time step, $dt$, to the embryo which had a mass of $M(t)$ at time $t$, where $\lambda = \textrm{ln}(M/M_{pl})/T_{grow}$ is the growth constant \citep{Dodson-Robinson08}.

The average eccentricity of Jupiter-mass exoplanets between 1.5 and 10 AU from their host star is 0.28\footnote{from the data compiled at www.exoplanet.eu as of July 2013}. Thus, the eccentricity of the embedded planet was varied from 0.0 to 0.3. A more eccentric planet would create a more dynamically excited and collisionally active planetesimal disk but this may result in a decrease in the planetesimal growth rate and loss of significant mass through collisional grinding. Therefore, it is initially unclear which orbital configuration produces more second-generation dust.

All simulations are in a gas-free disk for a number of reasons: 1) planetesimals in these simulations are all initially large ($R\sim 150$ km) so aerodynamic gas drag over the simulation time--scale is insignificant; 2) there is observational evidence that the inner hole of a transitional disk has low gas density \citep{Pontoppidan08, Brown08, Ireland08}; 3) it allows us to isolate the effects of collisional dynamics; 4) in this work dust is the instantaneous collisionally generated second generation dust, not primordial dust that is intimately mixed with the gas.

All simulations used a constant orbital (major) time-step of 0.01 years, providing good temporal resolution. However, minor steps were used in resolving the details of a planetesimal collision in the \emph{rubble} model.
\begin{figure*}[t]
	\centering
	\includegraphics[width=0.7\textwidth]{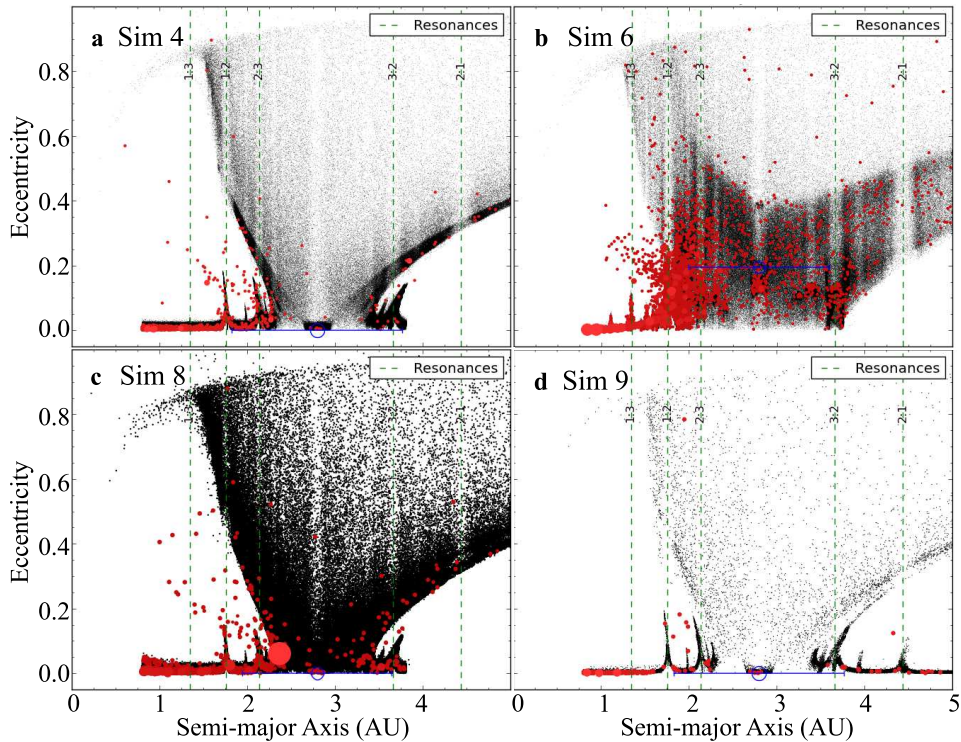}
	\caption{Planetesimals in eccentricity semi-major axis space. Size of markers is proportional to planetesimal mass. Black indicates $m<1.1m_0$, red $m>1.1m_0$, where $m$ is the mass of the planetesimal and $m_0$ is the initial mass of a planetesimal. A blue circle indicates the embedded planet with a radius of one Hill sphere ($R_H$). Error bars extend $5R_H$. Green dashed lines mark the location of major mean-motion resonances.}
	\label{fig:e0_0_e0_2_merge_rub_eVa}
\end{figure*}

\subsection{Planetesimal Collision Models}
\label{sec:collision_models}

Two different collision models were used: \emph{perfect merging} and \emph{rubble}.
\emph{Perfect merging} (sim. 0-6) treats all collisions as perfectly inelastic collisions and has been used in many numerical simulations due to its simplicity and speed \citep[i.e.][]{Brahic77,Kokubo96, Raymond11}. However, it does not allow fragmentation or erosion. This presents two problems: 1) collision outcomes may not be realistic; 2) an interpretive procedure is needed to determine where the highest concentration of the ``visible'' dust is located (see \S\ref{sec:image_dust_distrib}). Thus, in sim.~7 and 8 the \emph{rubble} model was used \citep{Leinhardt05, Leinhardt09}, which is more complex and realistic but computationally slow.

\begin{figure*}[t]
	\centering
	\includegraphics[width=.8\textwidth]{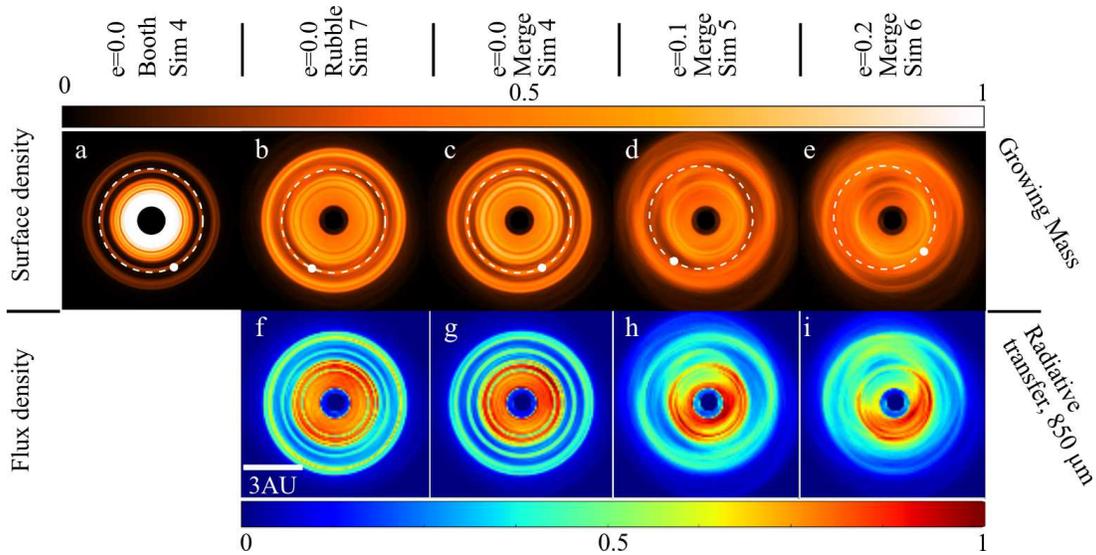}
	\caption{{\cone Top row c}alculated collisional dust surface density (see \S\ref{sec:image_dust_distrib}). Bottom row has surface brightness profiles in the $850 \mu m$ band for the simulations above them, created using the RADMC3D radiative transfer package \citep{RADMC3D}. {\cone Frame \textbf{a} shows sim.~4 if the dust surface density is tied directly to the planetesimal mass as described in \citet{Booth09}.}
Each frame is 10 AU square, the white dashed line is the orbit of the planet, the white dot is the location. The surface density frames are normalised to a common maximum, as are the flux density frames. The dust mass is normalised as it is impossible to precisely estimate it from these simulations. I.e.~The mass depends on the included size-range of dust, the model used when extrapolating dust mass (i.e. a power law as in (1), or a fixed ratio of dust mass to planetesimal mass), the number of particles in the simulation, and other unknown properties such as the tensile strength of the planetesimals. Therefore, the colour scale indicates \emph{relative} surface density.}
	\label{fig:fixed_grow_rub_dust_glow_radmc2}
\end{figure*}

\subsection{Calculating the Dust Distribution}
\label{sec:image_dust_distrib}
The \emph{rubble} model includes debris production. However, the \emph{perfect merging} collision model does not.
In previous work on debris disks \citet{Booth09} assumed that any collisional debris followed the mass. In his work each resolved planetesimal represented the massive end of a Dohnanyi collisional cascade \citep{Dohnanyi69}.
However, in our simulations the planetesimal disk is not in a steady state; instead it is rapidly evolving as the young solar system develops. Therefore, a Dohnanyi collisional cascade is not applicable. Instead we assume an instantaneous size distribution {\cone $n(r) dr = \zeta r^{-3.5} dr$, where $\zeta$ is a normalising factor, $n(r)$ is the number of objects with a radius between $r$ and $r+dr$, for the collision debris \citep{Leinhardt12}.}
The purpose of our algorithm is to obtain a realistic instantaneous spatial distribution of the collisional dust. This will depend upon the collisional environment of the planetesimals. Below we describe the algorithm:\\


{\noindent {\bf Step 1:} Determine the total mass crossing the orbit of each planetesimal, $M_{crossing}$. The orbits of two planetesimals intersect under the condition 
\begin{equation}
P_1 + P_2 \le \sqrt{A^2 + B^2},
\end{equation}
where $P$ is the orbital period, $A=e_2 P_1 cos(\psi_2) - e_1 P_2 cos(\psi_1)$, $B=e_2 P_1 sin(\psi_2) - e_1 P_2 sin(\psi_1)$, $\psi$ is the phase of the orbit.}\\

{\noindent {\bf Step 2:} Use the total mass and number of orbit crossing planetesimals ($N_{crossing}$) to calculate the average mass of a collider, 
\begin{equation}
m_{avg}=M_{crossing}/N_{crossing}.
\end{equation}
}\\

{\noindent {\bf Step 3:} Find the number of collisions necessary to encounter one target mass worth of colliders,
\begin{equation}
N_{coll} = m_{targ}/m_{avg},
\end{equation}
{\cone where $m_{targ}$ is the mass of the target body.}
}\\

{\noindent {\bf Step 4:} Scale the mass of a full size distribution of debris, $m_{smash}$, by the number of target masses encountered per orbit, 
\begin{equation}
\alpha_{coll} = \frac{P}{N_{coll} t_{coll}},
\end{equation}
where $t_{coll} = 1/(\chi v n_{avg})$ is the collisional timescale of the target, $\chi = \pi (R+r_{avg})^2 (1 + v_{esc}^2/\sigma_v^2)$ is the collisional cross-section of the target, $R$ is the target radius, $r_{avg}$ is the radius of $m_{avg}$, $v_{esc}$ the escape velocity from an object with a mass equal to the combined mass, $m_{targ}+m_{avg}$, $\sigma_v$ is the velocity dispersion of the orbit crossing planetesimals, $v$ is the speed of the target, and $n_{avg}$ is the number density of colliders.}\\

{\noindent {\bf Step 5:} Scale the mass of debris produced from the total disruption of the target, $m_{smash}$, by $\alpha_{coll}$ and calculate the mass of a single dust pixel, 
\begin{equation}
m_{dustpixel} = \alpha_{coll} m_{smash}/ N_{dustpixel},
\end{equation}
where $N_{dustpixel}$ is the number of dust units to be spread around the target's orbit (in this work $N_{dust pixel}=1000$).}\\

{\noindent {\bf Step 6:} Spread the dust around the target orbit in $\theta$ such that each dust pixel is spaced equally in time. Therefore,
\begin{equation}
d\theta = 2 \pi a b/(N_{dust pixel} r^2),
\end{equation}
where $a$ is the semi-major axis of the orbit, $b$ is the semi-minor axis of the orbit, and $r$ is the distance from the stellar focus to the dust pixel. }

\section{Results}	
The fixed--mass embedded planet simulations (sim. 0--3) are qualitatively similar to the corresponding growing--planet simulations (sim. 4--9) in all major respects (i.e. planetesimal, surface density, and dust distributions). However, the large impulse from the $1M_J$ planet at the start of the simulations reduces the disk particle number drastically, giving poor collision statistics. Therefore, only simulations with a growing planet (sim. 4--9) are discussed here.

Figure \ref{fig:e0_0_merge_surfD_evo} shows an example of the temporal evolution of the planetesimal surface density (in this case for sim.~4). The structures in the surface density evolution are qualitatively similar in both the equivalent \emph{rubble} simulation (sim.~7) and the fixed--planet mass simulation (sim.~0). All of the e=0.0 simulations result in a clearly defined gap around the orbit of the planet and additional partial gaps at low--order mean motion resonances. In addition there is a horseshoe ring of planetesimals that orbit with the planet in 1:1 resonance (visible from 330 yr). These obvious structures in the planetesimal disk become less clear with increasing planet eccentricity.

The collisionally active zones of the planetesimal disk are highlighted in Figure \ref{fig:e0_0_e0_2_merge_rub_eVa}. Significant planetesimal growth is generally confined to the inner planetesimal disk with more modest growth in the outer disk and the 1:1 resonance. In the low eccentricity simulations regions of growth are stimulated by low--eccentricity pileups such as those at the 1:2 and 2:3 inner resonances. However, it is clear from the extended planetesimal disk simulation (sim.~9, Fig.~\ref{fig:e0_0_e0_2_merge_rub_eVa}-\textbf{d}) that planetesimal growth is enhanced out to at least the outer 2:1 resonance. Unlike the smaller disk simulations (sim. 0--8) there are enough planetesimals at low eccentricity and larger semi-major axis to collisionally interact with the highly eccentric planetesimals that have been excited by the planet (those planetesimals on eccentric spurs -- sharp arc-like features). Increasing dust production both interior and exterior to the planet orbit.
In the high--eccentricity simulations, dynamical stirring increases the collision rate over a broad range in semi-major axis as in Figure.~\ref{fig:e0_0_e0_2_merge_rub_eVa}-\textbf{b}.
All simulations show eccentric spurs which increase the velocities and probabilities of collisions and are the main mechanism by which planetary stirring enhances collisional dust production.

Figure \ref{fig:fixed_grow_rub_dust_glow_radmc2}-\textbf{a} shows the surface density in dust if dust mass is tied directly to the planetesimal mass as is assumed in debris disks \citep{Booth09}. Figure \ref{fig:fixed_grow_rub_dust_glow_radmc2}-\textbf{b} to \textbf{e} use our new method described in \S\ref{sec:image_dust_distrib} which should be more realistic. If we assume that these disks are disruptive and in a quasi-steady state of collisional grinding then the inner disk will dominate the dust surface density. 
However, the reason that the inner disk is so bright in Figure \ref{fig:fixed_grow_rub_dust_glow_radmc2}-\textbf{a} is because the planetesimals at small semi-major axis are not significantly perturbed by the embedded planet and have grown to large masses. In the \citet{Booth09} method each planetesimal represents an entire collisional cascade, which generates significant dust mass. In reality the inner region of simulations is dynamically quiet and produces only a small amount of collisional debris because most collisions result in accretion events not disruption events. 
Thus, for the rest of this paper we will discuss only the instantaneously generated dust mass calculated with our algorithm in the context of young transitional objects. 

Simulations with a circular embedded planet (Fig.~\ref{fig:fixed_grow_rub_dust_glow_radmc2}-\textbf{b} and \textbf{c}) have a bright inner ring, dark gap with bright 1:1 resonance, and a bright outer ring. However, the simulations with an eccentric planet ($e>0.0$, Fig. \ref{fig:fixed_grow_rub_dust_glow_radmc2}-\textbf{d} and \textbf{e}) have less pronounced 1:1 resonances and an incomplete gap. The fixed--mass simulations 0--3 share these qualitative traits.
The murky gap in the eccentric simulations is at odds with the predictions from fluid simulations that predict a cleared eccentric gap \citep{Marzari10, Moeckel12}. However, the short time--scale of our simulations may be the cause of this discrepancy.

Flux density plots in the bottom row of Fig.~\ref{fig:fixed_grow_rub_dust_glow_radmc2} were created with RADMC3D \citep{RADMC3D}, a monte--carlo radiative transfer code, at a wavelength of $\lambda = 850 \mu m$. 
The total observed flux depends strongly on the dust mass, which in turn depends on the assumed disk mass. We do not make any attempt to predict flux but instead concentrate on the qualitative features expected in a disk that is initially optically thick.
Small dust was modelled as two dust species, amorphous carbon (AC) and silicates (SIL), both species having radii of $0.1 \mu m$ and $0.631 \mu m$, and relative abundances of 0.2 (AC) and 0.8 (SIL). Larger dust from $1 \mu m$ to $1000 \mu m$ is modelled using simple Mie scattering spheres with three size bins per decade. This range was chosen as dust of all sizes should be created in a collision. The smallest size is approximately defined by the dust blow--out radius, and the largest is when emission in the sub--mm is no longer significant. Features highlighted in the surface density plots are still present. In addition pericentre glow is visible in Figure \ref{fig:fixed_grow_rub_dust_glow_radmc2}-\textbf{h} and \textbf{i} along with an azimuthal asymmetry in the flux distribution of the inner disk.

\begin{figure*}
\centering
	\includegraphics[width=\textwidth]{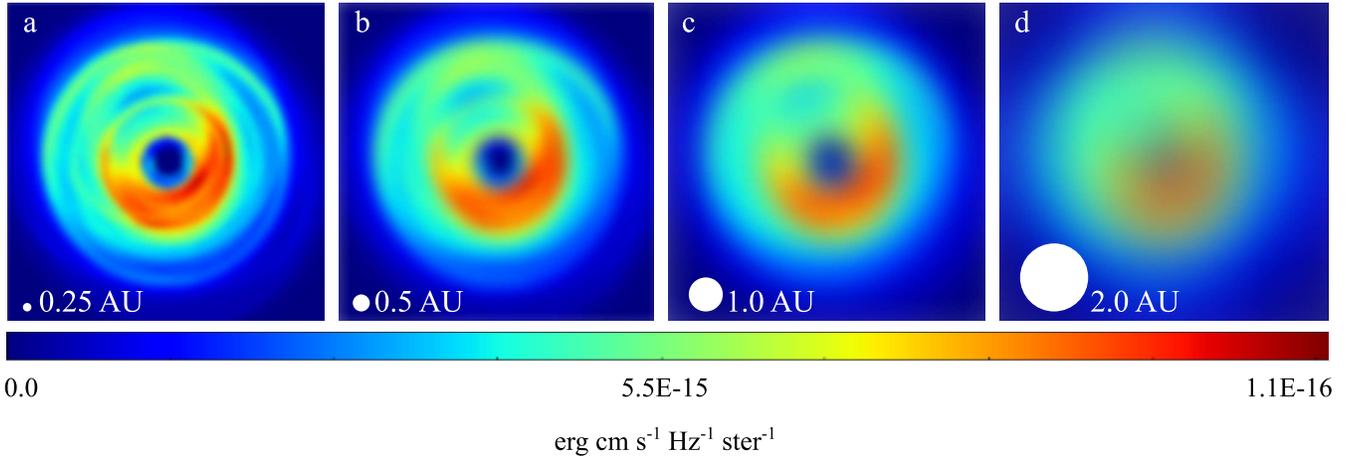}
	\caption{ Synthetic observations of sim.~6. Beam--size increases from left to right, white dot shows the beam--size to scale. The dust mass present in each frame is normalised to $2.8\times10^{-7} M_\odot$ (the mass of collisional dust produced in simulation 7 by the \emph{rubble} collision model) and is assumed to be observed straight on from 10 parsecs. The degradation of the image was performed by convolving the source image with a gaussian kernel which has a standard deviation equal to the displayed beam--size. The obvious asymmetry in frame \textbf{d} with only a minor eccentricity of the embedded planet shows that this could be a promising way to detect young planets.}
	\label{fig:beamsize}
\end{figure*}

Figure \ref{fig:beamsize} shows an illustrative example of the observational outcomes of our work. Assuming a dust mass of $2.8\times10^{-7} M_\odot$ (the mass of the debris produced in sim.~7) and the dust composition outlined above, the maximum signature is $\sim$200 mJy when observed face on at 10 parsecs. As this is derived from the surface density this depends upon: size--range of dust, extrapolation model from planetesimals to dust, and particle number. However, the disk gap would be visible up to a beam--size of $\sim$0.5 AU (0.05 '') and the asymmetry of the disk when e$>$0.0 up to a beam--size of $\sim$2 AU (0.2 '').

\section{Summary and Conclusions}
 In this work we present results from ten numerical simulations of collisional dust production from dynamical stirring by a young embedded planet. The simulations that we present use two different planetesimal collision models, \emph{perfect merging} and \emph{rubble}. Most simulations use a small 3AU-wide planetesimal disk and $10^6$ planetesimals, but we have also completed one lower resolution large disk simulation with a 10.2 AU-wide disk and $10^5$ planetesimals. 
A significant advance of this work is the use of a more realistic dust production model, increasing the accuracy of the spatial distribution of instantaneous collisionally produced dust and the resulting observables.

The series of simulations found that dust production is enhanced by the presence of a planetary companion. Dynamical stirring from the planet perturbs some planetesimals onto highly eccentric orbits (Fig.~\ref{fig:e0_0_e0_2_merge_rub_eVa}) which increases collisional activity from the inner 1:3 resonance to the outer 2:1 resonance producing two bright regions interior and exterior to the gap (Fig.~\ref{fig:fixed_grow_rub_dust_glow_radmc2}). The eccentric simulations show that a clean gap is not opened in contrast to the fluid case \citep{Marzari10, Moeckel12}, but the disk asymmetry can be easily observed (Fig.~\ref{fig:beamsize}).

\section{Further Work}	
{\cone The next steps will be to test and improve this model in numerous ways, observe the effect of multiple planets, improve the treatment of dust dynamics, and take account of the the dust lifetime.

The algorithm in \S\ref{sec:image_dust_distrib} neglects the effects of gas and primordial dust upon the dust dynamics. This work addresses the extreme case of no gas, but as the density of gas in the inner regions of transitional and pre--transitional disks is not well constrained \citep{Pontoppidan08, Brown08, Ireland08} other scenarios should be investigated.

The \emph{rubble} model is computationally intensive in a collisionally active disk. In the future  EDACM \citep{Leinhardt12} will be used. EDACM has the same fragmentation and debris tracking capabilities whilst using analytical expressions to determine outcomes, reducing computational load.

Finally, the exact magnitude and lifetime of the collisional dust cannot be determined from these simulations. Further investigation into this area will have to treat the dynamics and lifetime of collisional dust carefully.
}

\section{acknowledgements}
J.D. is grateful for support from NERC, Bristol School of Physics, and McDonald Observatory. Z.M.L. is supported by a STFC Advanced Fellowship.  S.D.R. is supported by National Science Foundation CAREER award AST-1055910. N.A.T. is supported by the Leverhulme Trust. Our thanks to Andy Young for help with data visualisation. This work was carried out using the computational facilities of the Advanced Computing Research Centre, University of Bristol\footnote{http://www.bris.ac.uk/acrc/}. We thank an anonymous referee for thoughtful comments that greatly improved the manuscript.



\end{document}